\documentclass[12pt]{article}
\usepackage{a4}
\usepackage{amsmath}
\usepackage{multirow}
\usepackage{graphicx}
\usepackage{rotating}
\usepackage{subfigure}
\numberwithin{equation}{section}
\newcommand{\beq}{\begin{equation}}
\newcommand{\eeq}{\end{equation}}
\begin{document}
\title{Spectrum of orientifold QCD in the strong coupling and hopping expansion approximation}
\author{Gregory Moraitis\\
\small{Physics Department, Swansea University, Singleton Park, Swansea SA2 8PP, UK}}
\date{}
\maketitle
\abstract{We use the strong coupling and hopping parameter expansions to calculate the pion and rho meson masses for lattice Yang-Mills gauge theories with fermions in irreducible two-index representations, namely the adjoint, symmetric and antisymmetric. The results are found to be consistent with orientifold planar equivalence, and leading order $1/N_c$ corrections are calculated in the lattice phase. An estimate of the critical bare mass, for which the pion is massless, is obtained as a function of the bare coupling. A comparison to data from the two-flavour SU(2) theory with adjoint fermions gives evidence for a bulk phase transition at $\beta_c\sim2$, separating a pure lattice phase from a phase smoothly connected to the continuum.}
\section{Introduction}
\label{sec:introduction}
Recently, there has been an ongoing effort to use lattice techniques for studying properties of gauge theories beyond QCD, with a large number of colours ($N_c\gg3$) and/or fermions in higher dimensional representations. The earliest motivation for this has been the proposed `orientifold planar equivalence' \cite{Armoni:2003gp, Armoni:2004ub} which is valid at infinite $N_c$, and provides a link between pairs of gauge theories (for a general review, see \cite{Armoni:2004uu}; for a lattice formulation, see \cite{Patella:2005vx}). As a particular case, it predicts that `adjoint QCD', `symmetric QCD' and `antisymmetric QCD' (by which we mean theories whose action is the same as QCD but with the fermions in the respective representation) all have the same bosonic sector at infinite $N_c$. We refer to these theories collectively as `orientifold theories'. It is interesting that by taking fermions in the adjoint, the one-flavour theory is identically $\mathcal N=1$ Super Yang-Mills, and so we can copy analytical predictions obtained using supersymmetry to the other two theories \cite{Armoni:2003fb}. Furthermore, for $N_c=3$, the antisymmetric theory becomes one-flavour fundamental QCD, suggesting a pathway for making real predictions in a close relative to real QCD -- provided $N_c=3$ is `close' to infinity. However, today this question can only be addressed by measuring the size of $1/N_c$ corrections non-perturbatively using lattice methods. In the case of pure Yang-Mills or quenched QCD with fundamental fermions, many studies have found that the corrections are indeed small (for a review, see \cite{Teper:2008yi}). There has been much less work on two-index fermions, though studies are now beginning to appear. In particular, in \cite{Armoni:2008nq}, a quenched lattice simulation of the quark condensate in orientifold theories was carried out, and a comparison with the analytic expression from \cite{Armoni:2003yv} supports the equivalence. Note however that for two-index fermions, the quenched theory and the dynamical theory are different at infinite $N_c$, so a definitive result has yet to appear.

Orientifold theories have also gained attention as candidates for Beyond the Standard Model physics, as their dynamics is potentially very different to QCD. In particular, there are proposals for `Technicolor' models of Dynamical Electro-Weak Symmetry Breaking where the Higgs is replaced by a composite bound state of strongly coupled higher-dimensional fermions. A recent concrete example making use of such a Higgs sector is Minimal Walking Technicolor \cite{Dietrich:2005jn}, which in our language is just SU(2) adjoint QCD with two flavours. Numerical lattice studies have already been carried out to determine the non-perturbative dynamics of this theory \cite{DelDebbio:2008zf,Catterall:2007yx,Catterall:2008qk,Catterall:2009sb, Hietanen:2008mr, Hietanen:2009az, DelDebbio:2009fd, Pica:2009hc, Bursa:2009we}, and there is mounting evidence of a conformal infra-red fixed point, or at least near-conformal behaviour -- a requirement for the walking scenario. There have also been numerical investigations of conformal behaviour in the case of SU(3) symmetric fermions \cite{Shamir:2008pb, DeGrand:2008kx, DeGrand:2009et, DeGrand:2009hu, Machtey:2009wu, Sinclair:2009ec}.

Since, however, unlike QCD, there is no experimental data to guide the interpretation of numerical results, it is important to learn as much as possible by analytical means. A weak-coupling analysis has already been performed \cite{DelDebbio:2008wb} and, among other conclusions, gives perturbative estimates for the ratio of $\Lambda$ parameters, $\Lambda_\mathrm{lat}/\Lambda_{\overline{MS}}$, and for the additive renormalisation of the quark mass and of fermion bilinears in this regime. The present paper is written in the same spirit, and looks at the opposite side of the lattice phase diagram by studying the meson spectrum in the strong coupling regime.

It is clear that a lattice strong coupling expansion has no relevance in the continuum limit, however our goal is to establish analytic results which, first, will give a starting point for choosing simulation parameters and, second, will provide information on the phase structure of the lattice theory. In particular, we derive formulae against which numerical data can be compared to ensure simulations are not in an `unphysical' phase (in the sense of not having a continuum limit).

In addition, the meson masses provide explicit observables for which to check orientifold planar equivalence at infinite $N_c$. Formally, a general proof of planar equivalence that holds to all orders in the strong coupling and hopping expansion has already been presented \cite{Patella:2005vx}, and the results of this paper should be considered a special case. The benefit of our direct calculation is that it provides explicit expressions for the meson masses at finite $N_c$.

Section \ref{sec:hopping} of the paper establishes the notation and describes the strategy. The calculation is performed in Section \ref{sec:masses}, closely following the diagrammatic method of \cite{Kawamoto:1980fd}. We enumerate the relevant diagrams and explain the generalisation from fundamental to two-index fermions, including a description of special cases which fall outside the general analysis. In Section \ref{sec:discussion} the results are discussed and a connection is made with available numerical data, finding two distinct regimes consistent with a lattice phase and a phase connected to the continuum. The conclusions are summarised in Section \ref{sec:conclusions}.

\section{Strong coupling and hopping expansion approximations}
\label{sec:hopping}
Here we set up the notation and outline the strategy. Discretising using Wilson fermions in lattice units ($a=1$), the action and Dirac operator is
\begin{eqnarray}
S & = & S_g + \sum_{x,y}\overline\psi(x)D(x,y)\psi(y)\\
S_g & = & -\frac{1}{g_0^2}\sum_{x,\mu>\nu} \mathrm{Tr}\left[U(x,\mu) U(x+\mu,\nu) U^\dagger(x+\nu,\mu) U^\dagger(x,\nu) + \mathrm{h.c.}\right] \nonumber\\
D(x,y) & = & \delta_{xy}-K(x,y)\nonumber\\
K(x,y) & = &  2\kappa \sum_{\mu} \left[ P^-_\mu V(x,\mu) \delta_{y,x+\mu} + P^+_\mu V(x-\mu,\mu)^\dagger \delta_{y,x-\mu}\right] \label{eq:hopping} \ ,
\end{eqnarray}
where $P_\mu^\pm=(1\pm\gamma_\mu)/2$ are the standard projectors, we have set the Wilson parameter $r$ to $1$, and the expansion parameter $\kappa$ is related to the bare quark mass by the usual relation $2\kappa=1/(4+m_0)$.We do not specify the number of flavours beyond stating $N_f\ge2$, as the final result will coincide with the quenched result to the order we work to in the hopping expansion. The gauge part of the action is the standard sum over elementary plaquettes, with the links $U$ always transforming in the fundamental representation of the gauge group. In the Dirac operator the links $V$ are in an arbitrary representation, which we will take to be either the fundamental or a two-index irreducible representation.

We want to compute the two-point meson correlator in the triplet channel,
\begin{eqnarray}
G_{\alpha\beta\gamma\delta}(x,y) & = & \langle\overline\psi_\alpha(x)\psi'_\beta(x)\overline{\psi'_\gamma}(y)\psi_\delta(y)\rangle\nonumber\\
 & = & -\frac 1Z\int \mathcal D[U](\mathrm{det}D)D^{-1}_{\alpha\delta}(x,y)D^{-1}_{\gamma\beta}(y,x) \mathrm e^{S_g} \ ,
\end{eqnarray}
where $\psi$ and $\psi'$ are different fermion flavours and colour indices are contracted to make colour singlet mesons. There is a well-established procedure for computing the correlator in the strong coupling/hopping expansion approximations:
\begin{enumerate}
\item Expand the quark propagators as a series in $\kappa$ (hopping expansion)\footnote{Note that this is a large mass expansion, immediately excluding using this approach for studying the open question of whether a theory is conformal in the IR.}. To each $\kappa$ is associated a factor given by one of the terms of \eqref{eq:hopping}, which geometrically corresponds to a link between two adjacent lattice sites. The full expansion is constructed by summing all possible discrete paths linking the two spacetime points $x$ and $y$. Note that backtracking paths are not allowed as we have made the choice $r=1$ for the Wilson parameter in the fermion action, so that $P^+_\mu P^-_\mu = 0$.
\item Expand the fermion determinant in powers of $\kappa^4$. This inserts an arbitrary number of closed four-link plaquettes (the product and trace of four factors \eqref{eq:hopping}) starting at any point on the lattice, in any orientation. There are $k$ (possibly overlapping) plaquettes at order $k$.
\item Expand the gauge part of the action in powers of $1/g_0^2$ (strong coupling expansion). This also produces four-link plaquettes as in the previous step, but containing only fundamental gauge links $U$ and no spin factors.
\item In the total expansion there will then be a mixture of links in differing representations ($U$ in the fundamental from the gauge part and $V$ in a two-index representation from the fermion part). The only non-zero contributions are those where the gauge links come in combinations which survive the Haar integrals. We will only need the two-link $U(N_c)$ integral,
\beq
\label{eq:orthogonality}
\int \mathcal D[U]R^a[U]_{ij}R^b[U^\dagger]_{kl}=\frac{1}{d_R}\delta^{ab}\delta_{il}\delta_{jk} \ ,
\eeq
where $R^a[U]$ denotes the link $U$ in representation $a$, and $d_R$ is the dimension of the representation. The difference between U($N_c$) and SU($N_c$) will be discussed in Section \ref{subsec:sun}.
\end{enumerate}

\begin{figure}[tbp]
\centering
\includegraphics{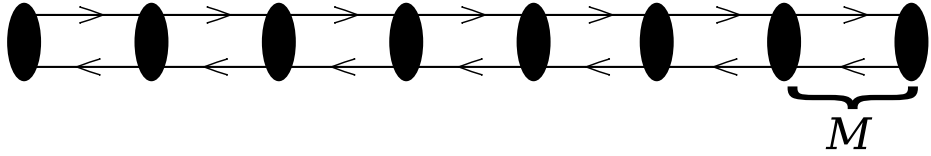}
\caption{In order for Haar integrals to be non-zero, links must be oppositely paired between each lattice site. If we set $e^{-S_g}\sim1$ and $\mathrm{det}D\sim1$, the only possibility is for the quark and antiquark propagators to be colinear.}
\label{fig:colinearpath}
\end{figure}

To make progress, we restrict ourselves to a fixed order in $\kappa$ and $1/g_0^2$. As a starting point, expand the gauge action $\mathrm e^{S_g}$ to lowest (zeroth) order in $1/g_0^2$, where there are no gauge plaquettes, and all gauge links have to come from the fermionic part of the action. This means that in order to satisfy \eqref{eq:orthogonality}, the two paths from two quark propagators must be colinear at all points (and in opposite directions)\footnote{There is also a possibility of non-colinear paths, with \eqref{eq:orthogonality} satisfied by insertions of plaquettes from the fermion determinant, but this is higher order in $\kappa$ than we consider here.}, see Figure \ref{fig:colinearpath}. The full path can be built up recursively from an elementary building block, $M$, which to this order is just two colinear links. We can then go to higher order by generalising $M$ to include non-colinear segments (i.e. to the next order, gauge squares, see Section \ref{subsec:nlo}). This method for constructing all paths was used in \cite{Kawamoto:1980fd} to compute the pion and rho masses in standard QCD. We refer to this paper for the technical details, and here only establish the notation.

Define $G_{L,\alpha\beta\gamma\delta}(n,0)$ to be the sum of all paths of length $L$ from the origin $0$ to $n$. The full correlator is then
\beq
\label{eq:prop}
G_{\alpha\beta\gamma\delta}(n,0)=\sum_{L=0}^{\infty}G_{L,\alpha\beta\gamma\delta}(n,0) \ .
\eeq
To calculate it, we can relate $G_L$ to $G_{L-1}$ through
\beq
\label{eq:recursion}
G_{L,\alpha\beta\gamma\delta}(n,0)=\sum_{n',\sigma,\tau}M_{\alpha\sigma\gamma\tau}(n,n')G_{L-1,\sigma\delta\tau\beta}(n',0) \ ,
\eeq
where $M$ contains the factors describing the propagation between sites $n$ and $n'$. It can then be shown (e.g. by looking at \eqref{eq:recursion} in Fourier space) that the full propagator is then just the sum of an infinite geometric series,
\beq
\label{eq:correlator}
G_{\alpha\beta\gamma\delta}(n,0)=\sum_{L=0}^{\infty}G_{L,\alpha\beta\gamma\delta}(n)=\left(\delta_{\alpha\delta}\delta_{\gamma\beta}-M_{\alpha\beta\gamma\delta}(n,0)\right)^{-1} \ .
\eeq
We can look at the desired channels by inserting the appropriate $\Gamma$ matrices,
\beq
G_{AB}(n,0)=\sum_{\alpha\beta\gamma\delta}(\Gamma_A)_{\alpha\beta}(\Gamma_B)_{\gamma\delta}G_{\alpha\beta\gamma\delta}(n,0) \ ,
\eeq
with the $\Gamma$ for each of the scalar, pseudoscalar, vector, axial vector and tensor channels given by
\begin{eqnarray}
\Gamma_S  =  \frac12 & \Gamma_P  =  \frac12\gamma_5  & \Gamma_{V_\mu}  =  \frac12\gamma_\mu \nonumber \\
\Gamma_{A_\mu}  =  \frac i2\gamma_\mu\gamma_5 &  & \Gamma_{T_{\mu\nu}}  =  \frac1{4i\sqrt2}\gamma_\nu\gamma_\mu  -  \gamma_\mu\gamma_\nu \ ,
\end{eqnarray}
and normalised as
\beq
\operatorname{Tr}\Gamma_A\Gamma_B=\delta_{AB}\equiv\left(1, 1, \delta_{\mu\nu},\delta_{\mu\nu},\frac12(\delta_{\alpha\mu}\delta_{\beta\nu}-\delta_{\alpha\nu}\delta_{\beta\mu})\right) \ .
\eeq
The masses are then extracted by looking in momentum space for poles in $G_{AB}$ (or, equivalently, zeros in its inverse) for a particle at rest with momentum $p_\mu=(im,\mathbf 0)$.
\section{Computing the masses}
\label{sec:masses}
\subsection{Leading order}
\label{subsec:leadingorder}
In the framework of section \ref{sec:hopping}, the problem reduces to writing an expression for $M(n,n')$. For the lowest order colinear paths, the expression is simple:
\begin{eqnarray}
\label{eq:Morder0}
M_{\alpha\beta\gamma\delta}(n,n') & = & (2\kappa)^2\sum_\mu[(P^-_\mu)_{\alpha\delta}(P^+_\mu)^T_{\gamma\beta}\delta_{n+\mu,n'}\nonumber\\
& & \qquad\qquad\qquad + (P^+_\mu)_{\alpha\delta}(P^-_\mu)^T_{\gamma\beta}\delta_{n-\mu,n'}] \ .
\end{eqnarray}
We have omitted the links $V$, as the colour contribution does not depend on the shape of the path and can be factorised and easily treated separately, giving a constant factor $d_R$. Substituting \eqref{eq:Morder0} into \eqref{eq:correlator} and taking the Fourier transform, we obtain the expression
\begin{eqnarray}
\label{eq:Gorder0}
G_{\alpha\beta\gamma\delta}(n,0) & = & -d_R\int_{-\pi}^\pi \frac{d^4p}{(2\pi)^4} \mathrm e^{ip_\mu}\tilde G_{\alpha\beta\gamma\delta}(p)\nonumber\\
\tilde G^{-1}_{\alpha\beta\gamma\delta}(p) & = & \delta_{\alpha\delta}\delta_{\gamma\beta} - (2\kappa)^2\sum_\mu[(P^-_\mu)_{\alpha\delta}(P^+_\mu)^T_{\gamma\beta}\mathrm e^{ip_\mu} \nonumber\\
& & \qquad\qquad\qquad\qquad +(P^+_\mu)_{\alpha\delta}(P^-_\mu)^T_{\gamma\beta}\mathrm e^{-ip_\mu}] \ .
\end{eqnarray}
Contracting \eqref{eq:Gorder0} with $\Gamma$ matrices then gives the momentum space correlator. Detailed expressions can be found in \cite{Kawamoto:1980fd}; we simply quote the final result after substituting $p_\mu=(im,\mathbf 0)$, for the pseudoscalar channel $\Gamma_P, \Gamma_{A_0}$ and the vector channel $\Gamma_{V_i}, \Gamma_{T_{0i}}$ (where $i$ is a spatial index).
\begin{eqnarray}
\tilde G^{-1}_{PA_0} & = & \left( \begin{array}{cc}
1-12\kappa^2-4\kappa^2\mathrm{cosh}m_\pi & 4i\kappa^2\mathrm{sinh}m_\pi \\
-4i\kappa^2\mathrm{sinh}m_\pi & 1-4\kappa^2\mathrm{cosh}m_\pi \end{array} \right)\\
\tilde G^{-1}_{V_iT_{0i}} & = & \left( \begin{array}{cc}
1-8\kappa^2-4\kappa^2\mathrm{cosh}m_\rho & 4i\sqrt\frac12\kappa^2\mathrm{sinh}m_\rho \\
-4i\sqrt\frac12\kappa^2\mathrm{sinh}m_\rho & \frac12(1-4\kappa^2-4\kappa^2\mathrm{cosh}m_\rho) \end{array} \right)
\end{eqnarray}
The zeros of these matrices, and thus the masses, can easily be found by solving $\mathrm{det}\tilde G^{-1}_{PA_0}=0$, $\mathrm{det}\tilde G^{-1}_{V_iT_{0i}}=0$, to give the well known results
\begin{eqnarray}
\mathrm{cosh}m_\pi  & = & 1+\frac{(1-16\kappa^2)(1-4\kappa^2)}{8\kappa^2(1-6\kappa^2)}\\
\mathrm{cosh}m_\rho & = & 1+\frac{(1-12\kappa^2)(1-8\kappa^2)}{8\kappa^2(1-6\kappa^2)}
\end{eqnarray}
where, in analogy to QCD, we call the pseudoscalar particle $\pi$ and the vector $\rho$. To this order, there is no dependence on representation.
\subsection{Next to leading order}
\label{subsec:nlo}
To write an expression for $M(n,n')$ at higher order one must jointly consider the $1/g_0^2$ and $\kappa$ expansion. We will choose to truncate at order $\kappa^6$, while the order for the strong coupling expansion will depend on the representation.

We consider the fundamental representation first. This case was already studied in \cite{Kawamoto:1980fd}, but it will be useful to present it to explain how to generalise to higher representations. When expanding $e^{S_g}$ to first order, we are allowing the placement of one gauge plaquette on the lattice. The edges must be paired with oppositely oriented links, and the only possible insertion is within a square of links from the fermion lines. $M(n,n')$ will therefore contain terms propagating from one site to the next not only colinearly but also leaving at most one empty square, to be filled by a plaquette from the strong coupling expansion. It is also possible to fill the square with a plaquette from the fermion determinant, but this introduces an extra factor of $\kappa^4$, which, together with at least four other fermion lines, would be higher order than we consider. There are a number of ways and orientations to construct this square, and they are all listed along with their order in Table \ref{tab:M}. The colour integral for each diagram can be evaluated, and all diagrams give a factor $1/g_0^2$ for the fundamental representation.

\begin{sidewaystable}[ptb]
\begin{center}
\begin{tabular}{|c|c|c|c|}
	\hline
\multicolumn{2}{|c|}{Order} & \multirow{2}{*}{Diagram} & \multirow{2}{*}{Spin contribution, including colour prefactor}  \\
\cline{1-2}
$1/g_0$ & $\kappa$ & & \\
	\hline
	0 & 2 & \includegraphics[scale=0.5]{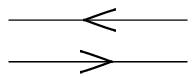} & $(2\kappa)^2(P^-_\mu)_{\alpha\beta}{(P^+_\mu)^T_{\gamma\delta}} $\\
	\hline
	$q$ & 4 & \includegraphics[scale=0.5]{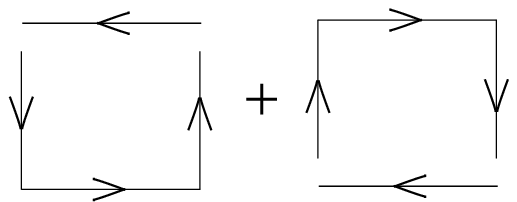} & $\frac{(2\kappa)^4}{4d_Rg_0^q}\sum_{\nu\neq\mu}\left[(P^+_\nu P^-_\mu P^-_\nu)_{\alpha\beta}(P^+_\mu)^T_{\gamma\delta} + (P^-_\mu)_{\alpha\beta}(P^-_\nu P^+_\mu P^+_\nu)_{\gamma\delta}\right] $ \\
	\hline
	$q$ & 6 & \includegraphics[scale=0.5]{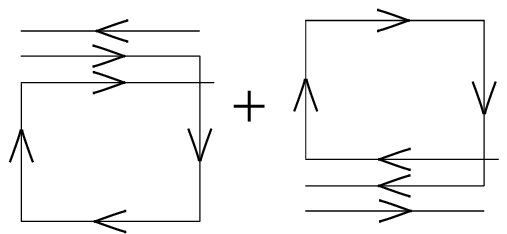} & $\frac{(2\kappa)^6}{4d_Rg_0^q}\sum_{\nu\neq\mu}\left[(P^-_\mu P^+_\nu P^+_\mu P^-_\nu P^-_\mu)_{\alpha\beta}{(P^+_\mu)^T_{\gamma\delta}} + (P^-_\mu)_{\alpha\beta}(P^+_\mu P^-_\nu P^-_\mu P^+_\nu P^+_\mu)^T_{\gamma\delta}\right] $ \\
	\hline
	$q$ & 6 & \includegraphics[scale=0.5]{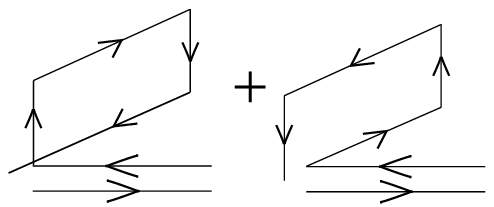} & $\frac{(2\kappa)^6}{4d_Rg_0^q}\sum_{\rho,\sigma\neq\mu}\left[(P^-_\mu)_{\alpha\beta}\left( (P^+_\mu P^-_\sigma P^-_\rho P^+_\sigma P^+_\rho)^T_{\gamma\delta} + (P^+_\mu P^-_\rho P^-_\sigma P^+_\rho P^+_\sigma)^T_{\gamma\delta}\right)\right] $ \\
        \hline
	$q$ & 6 & \includegraphics[scale=0.5]{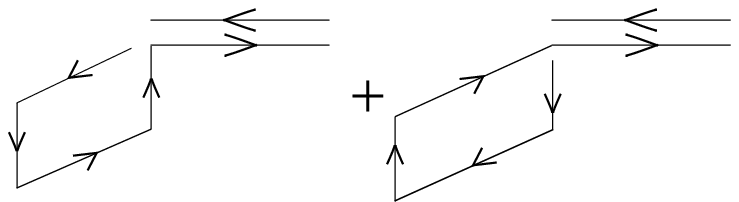} & $\frac{(2\kappa)^6}{4d_Rg_0^q}\sum_{\rho,\sigma\neq\mu}\left[\left( (P^-_\rho P^-_\sigma P^+_\rho P^+_\sigma P^-_\mu)_{\alpha\beta} + (P^-_\sigma P^-_\rho P^+_\sigma P^+_\rho P^-_\mu)_{\alpha\beta}\right)(P^+_\mu)^T_{\gamma\delta}\right] $ \\
	\hline
	$q$ & 4 & \includegraphics[scale=0.5]{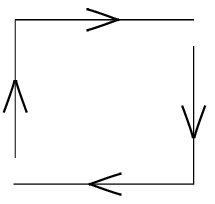} & $\frac{(2\kappa)^4}{4d_Rg_0^q}\sum_{\nu\neq\mu}\left[(P^-_\mu P^-_\nu)_{\alpha\beta}(P^+_\mu P^+_\nu)^T_{\gamma\delta}\right] $\\
	\hline
\end{tabular}
\end{center}
\caption{Diagrams contributing to $M(n,n')$. The first five rows represent diagrams linking sites $n$ to $n+\hat\mu$; there are analogous diagrams linking $n$ to $n-\hat\mu$. The last row links $n$ to $n+\hat\mu+\hat\nu$; there are also all permutations of terms with $\hat\mu\leftrightarrow-\hat\mu$ and $\hat\nu\leftrightarrow-\hat\nu$. In the fundamental representation $q=2$, while for two-index representations $q=4$.}
\label{tab:M}
\end{sidewaystable}

What happens if we change the representation of the fermions? The first thing to notice is that the diagrams as described above are all zero because of the orthogonality condition \eqref{eq:orthogonality}, since the fermions are in a two-index representation and the gauge links are in the fundamental. Since there are no other possibilities at order $1/g_0^2$, this whole order is zero and we must consider the next one. At second order, we can place two gauge plaquettes on the lattice. Overlapping them inside the fermion square (in the correct orientation, which is representation dependent) will yield a non-zero contribution.

For concreteness, consider the colour factor for the symmetric representation. The gauge integrals can be done by writing the reducible product of two overlapping fundamental links in terms of symmetric links $S$ and antisymmetric links $A$,
\beq
U_{ac}U_{bd}=S_{(ab),(cd)}+A_{[ab],[cd]} \,
\eeq
where $(ab)$ indicates the symmetric combination, with $b\geq a$, and $[ab]$ indicates the antisymmetric combination, with $b>a$. Defining a new basis such that $S_{(ab),(cd)}=S_{ij}E^i_{ab}E^j_{cd}$ (and similarly for the antisymmetric) with $i,j$ running from $1$ to $d_R$, the gauge integrals take the form
\beq
\int\mathcal D[U](S_{kl}+A_{kl})S^\dagger_{ij}
\eeq
By orthogonality, the second term is zero, and the first is given by \eqref{eq:orthogonality}. The argument is identical for antisymmetric fermion links, with the first term being zero and the second given by \eqref{eq:orthogonality}. Thus, we are back in a case exactly analogous to the fundamental, and the calculation follows through identically but with a different prefactor owing to the higher order expansion of $e^{S_G}$. There is a factor $2!$ from the Taylor expansion, and a combinatorial factor $1$ as there is exactly one way to place the two gauge plaquettes. This gives an overall colour factor of $1/(2g_0^4)$ for the symmetric and antisymmetric representations.

The case of the adjoint is similar, except for the fact that the two inserted gauge plaquettes must run in opposite orientations. There are now two ways to do this, so there is an extra factor of $2$ which cancels the $2!$ from the Taylor expansion. The result for the adjoint is thus $1/g_0^4$.

Having obtained the colour factor, the calculation for the spin part is the same for any representation, namely the same as the calculation in \cite{Kawamoto:1980fd}. Constructing $M$ from the diagrams in Table \ref{tab:M} and extracting the poles gives, for the pseudoscalar channel,
\begin{eqnarray}
\tilde G^{-1}_{PA_0} & = & \left( \begin{array}{cc}
H_{PP} & H_{PA_0}\\
H_{A_0P} & H_{A_0A_0}\\
\end{array} \right)\\
H_{PP} & = & 1-4\kappa^2[3+12\epsilon_R\kappa^2-96\epsilon_R\kappa^4\nonumber\\
 & & \qquad + (1+6\epsilon_R\kappa^2-24\epsilon_R\kappa^4)\mathrm{cosh}m_\pi]\nonumber\\
H_{PA_0} & = & -H_{A_0P}=4i\kappa^2(1+6\epsilon_R\kappa^2-24\epsilon_R\kappa^4)\mathrm{sinh}m_\pi\nonumber\\
H_{A_0A_0} & = & 1+4\kappa^2[3+9\epsilon_R\kappa^2-(1+6\epsilon_R\kappa^2+24\epsilon_R\kappa^4)\mathrm{cosh}m_\pi]\nonumber \ ,
\end{eqnarray}
and for the vector channel,
\begin{eqnarray}
\tilde G^{-1}_{V_iT_{0i}} & = & \left( \begin{array}{cc}
H_{V_iV_i} & H_{V_iT_{0i}}\\
H_{T_{0i}V_i} & H_{T_{0i}T_{i0}}\\
\end{array} \right)\\
H_{V_iV_j} & = & \delta_{ij}[1-4\kappa^2(2+6\epsilon_R\kappa^2-48\epsilon_R\kappa^4\nonumber\\ & & \qquad + (1+6\epsilon_R\kappa^2-24\epsilon_R\kappa^4)\mathrm{cosh}m_\rho)] \nonumber\\
H_{V_iT_{0j}} & = & H_{T_{0i}V_i} = -4i\delta_{ij}\frac12\kappa^2(1+6\epsilon_R\kappa^2-24\epsilon_R\kappa^4)\mathrm{sinh}m_\rho \nonumber\\
H_{T_{0i}T_{i0}} & = & \frac12[1-4\kappa^2+8\epsilon_R\kappa^4+96\epsilon_R\kappa^6\nonumber\\ & & \qquad - 4\kappa^2(1+6\epsilon_R\kappa^2-24\epsilon_R\kappa^4)\mathrm{cosh}m_\rho] \ , \nonumber
\end{eqnarray}
where the constant $\epsilon_R$ depends on the representation in the following way:
\begin{eqnarray}
\mathrm{fundamental}   \quad  \epsilon_R & = & \frac1{N_cg_0^2} \label{eq:efundamental}\\
\mathrm{adjoint}       \quad  \epsilon_R & = & \frac1{d_Rg_0^4}, \quad d_R=N_c^2-1\label{eq:eadjoint}\\
\mathrm{symmetric}     \quad  \epsilon_R & = & \frac1{2d_Rg_0^4}, \quad d_R=N_c(N_c+1)/2\label{eq:esymmetric}\\
\mathrm{antisymmetric} \quad  \epsilon_R & = & \frac1{2d_Rg_0^4}, \quad d_R=N_c(N_c-1)/2\label{eq:eantisymmetric}
\end{eqnarray}
Solving $\mathrm{det}\tilde G^{-1}_{PA_0}=0$, $\mathrm{det}\tilde G^{-1}_{V_iT_{0i}}=0$ gives the masses,
\begin{eqnarray}
\label{eq:masses}
\operatorname{cosh}m_\pi & = & 1+\frac{1-20\kappa^2+64\kappa^4-48\epsilon_R\kappa^4(1-8\kappa^2+64\kappa^4)}{8\kappa^2(1-6\kappa^2+6\epsilon_R\kappa^2(1-10\kappa^2+48\kappa^4))}\nonumber\\
\operatorname{cosh}m_\rho & = & 1+\frac{1-20\kappa^2+96\kappa^4-12\epsilon_R\kappa^4(5-84\kappa^2+384\kappa^4)}{8\kappa^2(1-6\kappa^2+6\epsilon_R\kappa^2(1-11\kappa^2+48\kappa^4))}
\end{eqnarray}
To this order, the meson correlators in the other channels do not contain poles leading to real-valued masses.
\subsection{U($N_c$) vs SU($N_c$)}
\label{subsec:sun}
\begin{figure}
\centering
\subfigure[Non-zero contributions to SU(2) symmetric and SU(4) antisymmetric. The diagrams are also present in SU(4) symmetric, but the one on the right vanishes.]
{
    \label{fig:sun:su2su4}
    \includegraphics[scale=0.9]{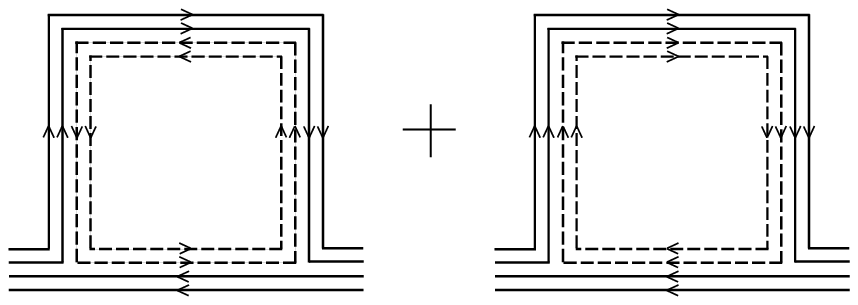}
}
\hspace{1cm}
\subfigure[This diagram for SU(3) symmetric vanishes due to symmetry properties of the representation.]
{
    \label{fig:sun:su3}
    \includegraphics[scale=0.9]{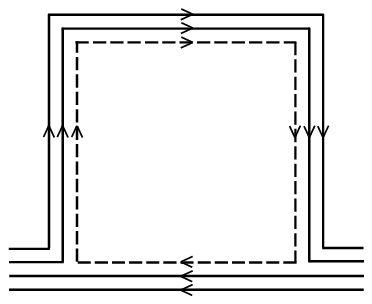}
}
\caption{Extra diagrams appearing in SU($N_c$), not present in U($N_c$). The solid lines are coming from the fermions, and are symmetrised or antisymmetrised. The dotted lines are insertions of plaquettes in the fundamental representation. All the diagrams which include squares in Table \ref{tab:M} have analogues to the above.}
\label{fig:sun}
\end{figure}

The masses as calculated in this section only make use of the U($N_c$) integral \eqref{eq:orthogonality}, so the results strictly apply only to U($N_c$) gauge theories, not SU($N_c$). If one is interested in the large-$N_c$ limit, this is not a problem as the singlet part of $\mathrm U(N_c) = \mathrm{SU}(N_c) \times \mathrm U(1)$ decouples, and both groups give the same result. However, for small gauge groups, one must also take into account contributions of the form
\beq
\label{eq:integral}
\int \mathcal D[U]U_{i_1j_1}U_{i_2j_2}\cdots U_{i_{N_c}j_{N_c}}=\frac{1}{N_c!}\epsilon_{i_1i_2\cdots i_{N_c}}\epsilon_{j_1j_2\cdots j_{N_c}} \ .
\eeq
With all links expressed in terms of the fundamental representation, the diagrams enumerated in Table \ref{tab:M} allow for up to four superimposed gauge links -- two from the two-index fermions and another two from the insertions of (up to) two plaquettes. Equation \eqref{eq:integral} is non-zero for $N_c$ superimposed links, so this integral will contribute for $N_c=2,3,4$. For $N_c\geq5$, to this order in the strong coupling expansion, SU($N_c$) coincides with U($N_c$). In addition, representations for some small gauge groups are equivalent to each other, and need to be considered as special cases. We look at each individually:
\begin{itemize}
\item SU($N_c$) adjoint --- For the adjoint representation, the full contribution is captured by including the two orientations of the gauge plaquettes, as we have done in Section \ref{subsec:nlo}, and equation \eqref{eq:integral} plays no role. The result \eqref{eq:eadjoint} is therefore unchanged for all $N_c$.
\item SU(2) antisymmteric --- This representation is just the singlet, so the theory is simply the free fermion theory. A strong coupling expansion is meaningless in this case, and we discard it completely.
\item SU(2) symmetric --- For SU(2), the symmetric and adjoint representations are unitarily equivalent. As the result for the adjoint has already been argued to be correct, the result for the symmetric must be the same. Alternatively, we can work directly in the symmetric representation, adding the two diagrams in Figure \ref{fig:sun:su2su4}. The two diagrams turn out to be equal, and this provides the factor of 2 needed to give $\epsilon_R=1/(d_Rg_0^4)$.  
\item SU(3) antisymmetric --- This representation is unitarily equivalent to SU(3) fundamental. The correct result must therefore be \eqref{eq:efundamental}, namely $\epsilon_R=1/(N_cg_0^2)$. The difference comes about because the orthogonality condition \eqref{eq:orthogonality} does not vanish to order $1/g_0^2$ as it does for the other two-index representations.
\item SU(3) symmetric --- From \eqref{eq:integral}, the only extra three-link diagram which could contribute is shown in Figure \ref{fig:sun:su3}. However, in this representation, the symmetrisation of the indices leads to the vanishing of the diagram. Thus \eqref{eq:esymmetric} is correct without modification for SU(3).
\item SU(4) antisymmetric --- This case is similar to SU(2) symmetric: the same two diagrams in Figure \ref{fig:sun:su2su4} contribute and they are both equal, giving an extra factor of 2 compared to U(4). The result is $\epsilon_R=1/d_Rg_0^4$. 
\item SU(4) symmetric --- As for SU(3) symmetric, the symmetrisation of the indices make the diagram on the right in Figure \ref{fig:sun:su2su4} vanish, so \eqref{eq:esymmetric} is valid as it stands.
\end{itemize}
\section{Discussion}
\label{sec:discussion}
\begin{figure}[ptb]
\centering
\includegraphics[scale=0.4]{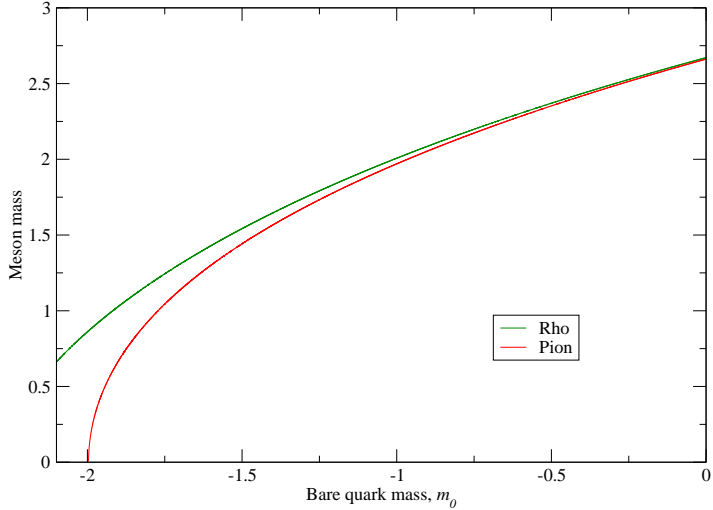}
\caption{Pion and rho mass as a function of the bare quark mass (adjoint representation). Parameters: $N_c=6$, $\beta=10.0$, with beta defined in the usual way, $\beta=2N_c/g_0^2$. The choice $N_c=6$ was chosen to keep away from the special cases discussed in Section \ref{subsec:sun}.}
\label{fig:masses:pion-rho}
\end{figure}

\begin{figure}[pbt]
\centering
\includegraphics[scale=0.4]{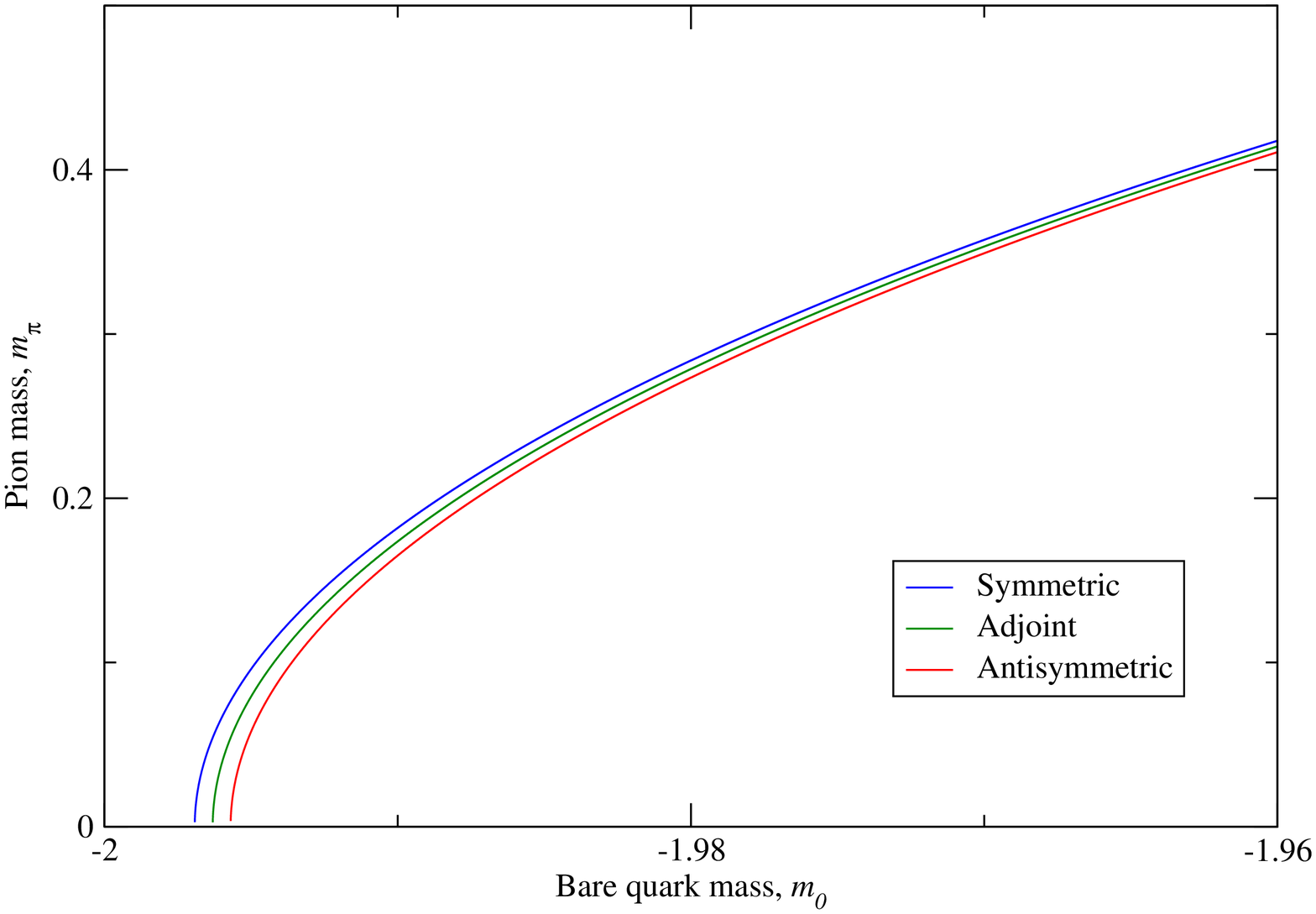}
\caption{Pion mass as a function of the bare quark mass for the three irreducible two-index representations. Parameters: $N_c=6$, $\beta=5.0$.}
\label{fig:masses:representations}
\end{figure}

The equations \eqref{eq:masses} give analytical predictions for the pion and rho masses in the strong coupling and hopping parameter expansions. We can use them to make a few general observations which may help in future lattice studies of orientifold theories.
The critical value of $\kappa$ where the pion vanishes can be calculated as a function of the bare coupling and is found to be
\beq
\label{eq:kappac}
\kappa_c \approx \frac14\left(1-\frac{3}{32}\epsilon_R\right) \ .
\eeq
Thus, moving away from the infinite coupling limit has the effect of reducing $\kappa_c$ below $1/4$ (although the first order correction is very small). Note also that at $\kappa_c$ the pion mass is zero but the rho mass remains finite. Indeed, the rho mass is always above the pion mass (Figure \ref{fig:masses:pion-rho}), and the strong coupling phase is qualitatively similar to fundamental QCD. Furthermore, at any given $\kappa$, there is a definite ordering in the masses (both for the pion and the rho); the different factors $\epsilon_R$  are such that the symmetric is always heaviest, followed by the adjoint and then the antisymmetric (Figure \ref{fig:masses:representations}). While this has no physical significance, it is relevant for numerical simulations, as it tells us that the values of $\kappa$ needed to approach the chiral limit will be similarly ordered, with symmetric highest and antisymmetric lowest. 

\begin{figure}[tbp]
\centering
\includegraphics[scale=0.4]{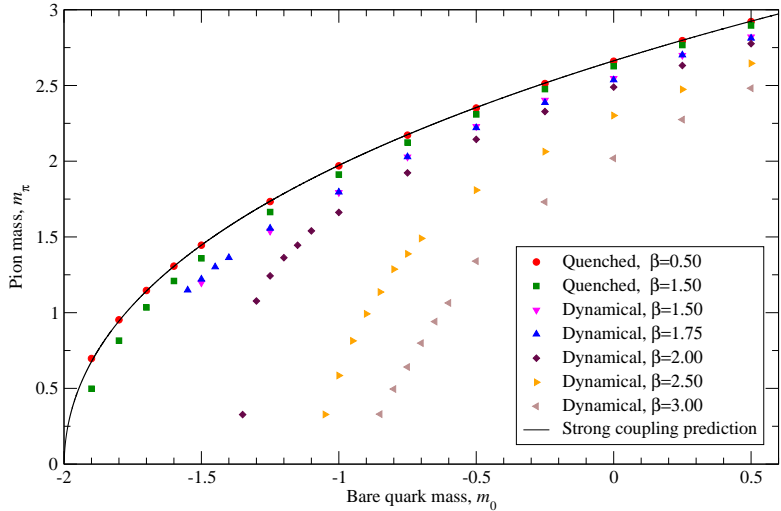}
\caption{Comparison of numerical lattice data with strong coupling prediction. The quenched data is our own, while the dynamical has been taken from \cite{Catterall:2008qk}. The strong coupling curve is plotted for $\beta=0.5$, but in practice shifts very little in the range $\beta=0.5-3.0$.}
\label{fig:comparison}
\end{figure}

It is interesting to compare \eqref{eq:masses} with numerical lattice data. Figure \ref{fig:comparison} plots the pion mass for the SU(2) adjoint theory, using data from \cite{Catterall:2008qk} for two-flavour dynamical simulations supplemented by our own quenched simulations. The strong coupling line is plotted for $\beta=0.5$, but in practice does not move significantly in the range $\beta=0.5-3.0$. For $\beta=0.5$, deep in the strong coupling phase, the lattice data falls on top of the strong coupling prediction (note that this is not a fit as \eqref{eq:masses} has no free parameters). Increasing to $\beta=1.5$, still quenched, leads to a small deviation, more notable for small masses, which is likely to be explained by going to higher order in the hopping expansion. At large masses, the weak dependence on $g_0$ seen in \eqref{eq:kappac} is borne out in this phase. The dynamical simulations are slightly puzzling: for $\beta=1.5$, one would expect the quenched and dynamical results to coincide for large masses, as the effects of the fermion determinant become negligible. This is not observed, suggesting either that the mass is simply not large enough, or that there could be a small underestimated systematic error in the data. With this uncertainty in mind, we can say that the data for $0.5<\beta<1.75$ is good agreement with the strong coupling prediction. In contrast, for $\beta\ge2$, there are significant departures from strong coupling, both in the magnitude of the masses and even in the qualitative behaviour as one approaches light quark masses. This is consistent with the finding in \cite{Catterall:2008qk} that there is a bulk phase transition at $\beta_c\sim2$, with a strong coupling lattice phase possessing no continuum limit for $\beta<\beta_c$, and a phase smoothly connected to the continuum for $\beta>\beta_c$.

The results are also consistent with orientifold planar equivalence, as the factors $\epsilon_R$ for the two-index representations all tend to the same value as $N_c\to\infty$ (notice that there is a cancellation in $\epsilon_R$ between factors of $2$ of different origin, coming together to ensure the equivalence works). In the large-$N_c$ limit, we have the asymptotic forms of \eqref{eq:masses},
\begin{eqnarray}
m_\pi & = & \mathrm{acosh}\left(\frac{1-12\kappa^2+16\kappa^4}{8\kappa^2(1-6\kappa^2)}\right) - 6\epsilon_R\kappa^2\nonumber\\
m_\rho & = & \mathrm{acosh}\left(\frac{1-12\kappa^2+48\kappa^4}{8\kappa^2(1-6\kappa^2)}\right) - 6\epsilon_R\kappa^2 \ .
\end{eqnarray}
Note, however, that this leading correction to the large-$N_c$ limit is not expected to be universal, and may be specific to the regime of validity of the expansion: the lattice strong coupling and large mass phase.
\section{Conclusions}
\label{sec:conclusions}
We have computed analytic expressions for masses of the $\pi$ and the $\rho$ mesons in the strong coupling and hopping expansion (large mass) approximations, for fermions in the three irreducible two-index representations. In the limit $N_c\to\infty$, the three converge to the same value, as predicted by the formal proof of orientifold planar equivalence on the lattice presented in \cite{Patella:2005vx}. In addition, we have extracted the leading $1/N_c$ corrections, which in the strong coupling phase are expressed only in terms of the dimensionality of the representation, $d_R$ (and the bare quark mass).

The results are already useful in understanding the lattice phase structure emerging from Monte-Carlo simulations. By comparing with recent numerical determinations of meson masses in two-flavour SU(2) adjoint QCD, we find evidence of two phases, supporting the conclusions of \cite{Catterall:2008qk}. The theory has a bulk phase transition with a strong coupling lattice phase, having no continuum limit, on one side ($\beta<2$), and a phase smoothly connected to the continuum on the other ($\beta>2$). As more simulations are performed for as yet unstudied theories, it is our hope that the results of this paper will help in recognising the phase structure and locating the correct region of parameter space to use for extracting continuum physics.

\section*{Acknowledgments}
I am greatly indebted to A. Patella and B. Lucini for their insights and careful reading of the manuscript, and to A. Armoni for numerous discussions and thoughtful analysis. I would also like to thank M. L\"uscher and the CERN theory department, where the bulk of this work was carried out.

\bibliography{strongcoupling}

\end{document}